\documentstyle[12pt,aaspp]{article}
\def\epsvec{\hbox{\bf e}}
\def\Ginga{\hbox{\it Ginga}}
\tighten
\received{1994 January 21}
\revised{DATE}
\accepted{1994 April 29}
\journalid{VOL}{JOURNAL DATE}
\articleid{Start}{Finish}
\paperid{Manuscript id}
\lefthead{BAND}
\righthead{BATSE GRB Line Search: II. Bayesian Consistency}

\begin{document}
\title{BATSE GAMMA-RAY BURST LINE SEARCH: \\
II. BAYESIAN CONSISTENCY METHODOLOGY}
\author{D. L. Band, L. A. Ford, J. L. Matteson}
\affil{CASS, University of California, San Diego, CA  92093}
\author{M. Briggs, W. Paciesas, G. Pendleton, R. Preece}
\affil{University of Alabama at Huntsville, Huntsville, AL 35899}
\author{D. Palmer, B. Teegarden, B. Schaefer}
\affil{NASA/GSFC, Code 661, Greenbelt, MD 20770}
\centerline{\it Received: 1994 January 21; accepted: 1994 April 29}
\centerline{\it To appear in the October 20 issue of the Astrophysical Journal}
\begin{abstract}
We describe a Bayesian methodology to evaluate the consistency between
the reported \Ginga\ and BATSE detections of absorption features in gamma
ray burst spectra.  Currently no features have been detected by BATSE, but
this methodology will still be applicable if and when such features are
discovered.  The Bayesian methodology permits the comparison of hypotheses
regarding the two detectors' observations, and makes explicit the subjective
aspects of our analysis (e.g., the quantification of our confidence in
detector performance).  We also present non-Bayesian consistency statistics.
Based on preliminary calculations of line detectability we find that both
the Bayesian and non-Bayesian techniques show that the BATSE and \Ginga\
observations are consistent given our understanding of these detectors.
\end{abstract}
\section{INTRODUCTION}
The presence or absence of absorption lines in the gamma ray burst (GRB)
spectra observed by the detectors of the Burst and Transient Source
Experiment (BATSE) on board the {\it Compton Gamma Ray Observatory (GRO)} is
one of the most pressing issues in the BATSE study of GRBs.  The absorption
lines observed in the 15-75~keV band by earlier GRB instruments
(Konus---Mazets et al. 1981; {\it HEAO-1}---Hueter 1987; \Ginga---Murakami
et al. 1988) were interpreted as cyclotron absorption in a teragauss
magnetic field (e.g., Wang et al. 1989), and as such, reinforced the
identification of the GRB sources with Galactic neutron stars.  Neutron stars
are the only known astrophysical site of such strong fields.  However, the
observed locations and intensities of the BATSE bursts (Meegan et al. 1992)
have undermined the Galactic neutron star paradigm for burst origin.  The
angular
distribution is isotropic, yet the intensity distribution is inconsistent
with a uniform, three-dimensional Euclidean source density.  Therefore, we
are at the center of a spherical source distribution which decreases radially,
and not within a disk population.  Consequently the search for absorption
features in the BATSE spectra has taken on additional importance.

No definitive lines have been discovered by the BATSE team thus far.  We have
reported this in shorter, less complete presentations (Teegarden et al. 1993;
Band et al. 1993a; Palmer et al. 1994a); in the current series of papers,
beginning with Palmer (1994b), we
describe our search and analysis methods in greater detail.  Since this is an
ongoing search our analysis methods and results will undoubtably change over
the course of this series of publications.

In the current report we present the methodology by which we compare the
\Ginga\ and BATSE detections and nondetections.  While the presentation of
this methodology is our primary objective here, we demonstrate the use of the
resulting formulae with an approximation to the observed data, and therefore
we draw relevant conclusions about BATSE-\Ginga\ consistency which will most
likely remain true for more accurate calculations.  We derive the consistency
statistics using a Bayesian formulation; however, the meaning of the resulting
expressions does not require a detailed understanding of Bayesian methodology.
Built on a Bayesian foundation, this analysis employs concepts which are
formally not permitted in classical ``frequentist'' statistics (e.g.,
distributions for unknown parameter values, not just ``random'' variables), but
nonetheless the derived formulae should be considered reasonable and therefore
acceptable to the astrophysical community.
In our derivation we describe the Bayesian concepts where they are applied,
and note deviations from orthodox frequentist or Bayesian usage.  We also
present frequentist consistency statistics which we find lead to similar
conclusions as the Bayesian formulae.

One of the central tenants of Bayesian inference is that probabilities are
measures of our confidence in the truth of propositions, rather than simply
the frequency with which a result occurs (e.g., Loredo 1990).  Therefore
the probability that a hypothesis is true can be evaluated based on both
prior quantitative information (and more qualitative expectations) and new
observations.  This permits hypotheses to be compared by comparing
their probabilities.  In our case we ask:
given the observations, what are the odds---the ratio of the hypothesis
probabilities---that the BATSE and \Ginga\ results are consistent compared
to inconsistent?  Below we will explicitly compare the consistency hypothesis
(represented as proposition $H_0$) to various alternative hypotheses
(propositions $H_x$); for example, $H_1$ states
that because of a detector defect, BATSE is unable to detect the absorption
lines which are present.  The observations may be very unlikely for
both hypotheses (e.g., for results drawn from a continuum of possibilities),
yet can still favor one over the other.

The Bayesian formulation has a number of virtues.  First, this approach
permits us to frame the consistency calculation in terms of the observed
distribution of detections and nondetections, and does not require assumptions
about the population of results from which the observations may have been
drawn.  Second, while Bayesian inference has frequently been criticized for
the apparent arbitrariness in quantifying prior expectations, this merely
makes explicit an arbitrariness that is also present in ``frequentist''
methods.  For example, the threshold which a frequentist statistic must
exceed before we accept a conclusion is based on our expectation as to the
likelihood of that conclusion:  a conclusion contrary to our expectations
requires a more extreme threshold.  Third, the Bayesian
formulation provides guidance as to the elimination of unknown parameters
whose values are necessary for deciding the consistency question
but are not intrinsically interesting for this issue
(i.e., ``marginalization'' of ``nuisance'' parameters).  In our case the
frequency with which lines occur is the nuisance parameter.  Finally,
this methodology will continue
to be applicable if and when we do detect cyclotron lines in BATSE bursts
since it does not depend on any particular pattern of detections and
nondetections.

While absorption features have been reported by a number of instruments,
for the purpose of quantitative comparison we need well documented
details of both the detections and the bursts which were searched.
Similarly, we consider only the statistically significant lines
reported by other detectors.
We evaluate the line significance with the F-test which compares
fits to the spectrum of a continuum with and without lines.  The
F-test gives the probability that the decrease in $\chi^2$ of the
continuum plus line model versus the continuum model alone is
due to chance when no line actually exists
(Martin 1971, pp.~144-147), in other words, the probability
that a fluctuation of the continuum would appear as
significant as the observed feature.  Note that a
smaller probability indicates a more significant line.
For the BATSE detectors we have established a detection threshold
of an F-test probability less than $P_F<10^{-4}$; the threshold has
been chosen to eliminate spurious detections.  Also, the spectra
obtained by all detectors capable of observing the feature must be
consistent.  As we will show, currently the BATSE observations can
be compared to only two \Ginga\ detections.

Absorption features were reported to be present in $\sim 20$\% of the Konus
bursts (Mazets et al. 1981).  However the Konus spectra were analyzed
under the assumption that the continuum was $N_E\propto E^{-1} \exp(-E/E_0)$
(Mazets et al. 1982, 1983) while we find that low energy GRB spectra are best
fit by a variety of predominantly flatter spectral models (Band et al. 1993b);
we do not know whether the reported lines would be significant with a more
realistic continuum.  Finally, the significance of the line features and their
detectability (the probabilities for actual and spurious detections) in all
the Konus bursts have not been provided, precluding quantitative comparison.

Two line detections have been reported among the 21 {\it HEAO-1} bursts
(Hueter 1987), one with a significance of $5.6\times 10^{-4}$ and the other
with a significance of $3\times 10^{-3}$.  Neither of these line candidates
would qualify as detections by our detection threshold.  In addition, we have
no information about line detectability in the ensemble of {\it HEAO-1} bursts.

The \Ginga\ bursts provide the best documented detections.  Four sets of
lines in the \Ginga\ bursts have been reported, but only two sets meet
our detection criterion of $P_F < 10^{-4}$ (the following line significances
have been recalculated
using the fit parameters in the indicated references). Thus the lines in the
S2 segment of GB870303 (with a significance of $1.1 \times 10^{-3}$---Murakami
et al. 1988) and in GB890929 (a significance of $2.7\times 10^{-3}$---Yoshida
et al. 1991) cannot be considered detections.  The harmonically spaced lines
at 19.3 and 38.6~keV in GB880205 ($2.4\times 10^{-5}$---Murakami et al. 1988),
and the single line at 21.1~keV in the S1 segment of GB870303 ($1.5 \times
10^{-7}$---Graziani et al. 1992) constitute the Ginga detections.  Although
the line in GB870303 is formally very significant, the low
signal-to-background of the continuum and the small final $\chi^2$ (14.49 for
30 degrees-of-freedom, $P(\chi^2<14.49)=7.5\times 10^{-3}$) make this feature
suspect.

Therefore we compare the BATSE nondetections to the \Ginga\ detections.  In
the absence of a large enough ensemble of detections to characterize the
distribution of line parameters, we use the two \Ginga\ line detections to
define two line types.  In our example we calculate quantities for each
\Ginga\ line type separately, and also for the two types together.

Previously (Band et al. 1993c) we defined the consistency statistic as the
(non-Bayesian) probability of two or more detections in any of the \Ginga\
bursts and none in the BATSE data, that is, of a result at least as discrepant
as the observations.  This probability is a function of the unknown line
frequencies.  Maximizing this statistic with respect to the line frequencies
placed an upper limit on this probability of $\sim 5$\%.
However, this probability tells us how
unlikely the pattern of detections and nondetections is, but does not directly
inform us whether there is an inconsistency.  As we discuss below, the
observations may be even more unlikely under various hypotheses
about inconsistencies between the two instruments.  This was the motivation
for the adoption of a Bayesian analysis.  Nonetheless, we also present
non-Bayesian consistency statistics which we find lead to similar conclusions,
at least for our example based on approximations to the data.

Because of the small number of detections, we derive a likelihood function
for discrete line types and only comment briefly on continuous
distributions of line parameters (\S 2.1).  This likelihood function is
required by the Bayesian methodology (\S 2.2); the Bayesian formalism also
provides estimates of the line frequencies which are both interesting in their
own right and useful for the consistency analysis.
Using this methodology we compare the consistency hypothesis to
various hypotheses about possible sources of the apparent discrepancy between
detectors (\S 3); these formulae are then applied to an illustrative example
which approximates the observations (\S 4).  For completeness we present our
earlier frequentist consistency calculations (\S 5).  The implications of
these various consistency measures are discussed in \S 6, after which we
summarize our conclusions (\S 7).  We use the standard notation where
$p(a \,|\, b)$ means the probability of proposition $a$ given proposition $b$
(a proposition may be a hypothesis, a model's validity or its parameter list).
\section{BAYESIAN CONSISTENCY PROBABILITY}
\subsection{Likelihood Function}
The population of absorption features is characterized by distributions of
line parameters such as energy centroids, equivalent widths, intrinsic widths,
harmonics, etc.  Undoubtably these parameters vary continuously.  However,
the number of detections is insufficient to determine these distributions.
Instead of attempting to model the parameter distributions, we restrict the
line population to the two line types defined by the \Ginga\ detections.
Therefore we develop our methodology for a finite number of line types,
and only comment briefly on the continuum limit.  First we develop the
likelihood function, the probability of obtaining the data under a set of
hypotheses (which include our understanding of the instruments), and then
we embed this likelihood in a Bayesian framework.

Assume there are $n_t$ line types, each defined by a set of parameters
$\epsvec_\rho$, where $\rho$ denotes the line type (1 or 2 for the line
population based on the \Ginga\ detections).  Let $f_\rho=f(\epsvec_\rho)$
be the frequency with which line type $\rho$ (defined by $\epsvec_\rho$)
occurs in bursts, regardless of whether the line is detectable or whether
other line types are present.  In the absence of information about different
burst populations,
we assume the entire burst population is characterized by the same
line type distribution $f=f(\epsvec)$ (i.e., $f$ is the set of all $f_\rho$);
thus we do not assume that lines are evident only in long duration multispike
bursts, for example.  Note that we allow the
existence of more than one line type in a burst.  This is justified by the
presence of the S1 line at $\sim 20$~keV and the S2 lines at $\sim 20$ and
$\sim 40$~keV in GB870303 (although the second set of lines
is {\it not} significant enough to be considered a detection).  Further we
postulate that each line type is independent of the presence of all other
line types.

We represent the detection or nondetection of the line with parameters
$\epsvec_\rho$ in the $i$th burst by the propositions
$L_i(\epsvec_\rho)$ and $\bar L_i (\epsvec_\rho)$, respectively, and the
existence or absence of the line by $l_i(\epsvec_\rho)$ and
$\bar l_i(\epsvec_\rho)$, respectively.  Our assessment of the detection
probabilities depends on
our understanding of detector responses, etc.  We make this dependence
explicit by including the proposition $I$, representing our knowledge of the
observations, detector performance, etc., as one of the conditions in our
expressions.  Similarly, the probabilities will depend on the hypothesis
$H$ which we are evaluating.  Finally, the probabilities are functions of the
line frequency distribution $f$, which must be modeled if unknown
(as is currently the case).

Therefore we can express the probability of detecting a line as
\begin{eqnarray}
p \left(L_i(\epsvec_\rho) \,|\, fHI\right) &=&
   p\left(L_i(\epsvec_\rho) \,|\, l_i(\epsvec_\rho) fHI\right)
   p\left(l_i(\epsvec_\rho) \,|\, fHI\right) \\
   &+&
   p\left(L_i(\epsvec_\rho) \,|\, \bar l_i(\epsvec_\rho) fHI\right)
   p\left(\bar l_i(\epsvec_\rho) \,|\, fHI\right) \quad . \nonumber
\end{eqnarray}
The first term on the right is the probability for detecting real lines, while
the second term is the probability for a false positive.  We assume that the
line types are distinct enough that one cannot be confused with another,
otherwise
$p(L(\epsvec_\rho)\,|\,l(\epsvec_\rho) fHI)p(l(\epsvec_\rho)\,|\,fHI)$
must be replaced by
$\sum_{\sigma=1}^{n_t} p(L(\epsvec_\rho)\,|\,l(\epsvec_\sigma) fHI)
p(l(\epsvec_\sigma)\,|\,fHI)$.  Clearly if we include more line types which
are similar to each other this assumption that line confusion can be ignored
is less justified.  Since $l_i(\epsvec_\rho)$
and $\bar l_i(\epsvec_\rho)$ on the one hand, and $L_i(\epsvec_\rho)$ and
$\bar L_i(\epsvec_\rho)$ on the other, are exhaustive (i.e., the sum of their
probabilities equals 1),
\begin{eqnarray}
p\left(L_i\left(\epsvec_\rho\right) \,|\, fHI\right) &=&
   \alpha_{i\rho} f_\rho + \beta_{i\rho} (1-f_\rho) \\
p\left(\bar L_i(\epsvec_\rho) \,|\, fHI\right) &=&
   1- p\left(L_i\left(\epsvec_\rho\right) \,|\, fHI\right) =
   \left(1-\alpha_{i\rho}\right) f_\rho +
   \left(1-\beta_{i\rho}\right)(1-f_\rho) \quad , \nonumber
\end{eqnarray}
where $\alpha_{i\rho}$, the detection probability, and
$\beta_{i\rho}$, the probability of a spurious detection, must be
calculated specifically for the $i$th burst and $\rho$th line type, and may
depend on the hypothesis $H$ under evaluation.  We postpone the description of
how we calculate $\alpha$ and $\beta$ to a later publication in this series.

We are presented with a pattern of detections and nondetections of absorption
lines in the BATSE and \Ginga\ data, an observed realization from among all
possible outcomes, which we denote by the proposition $D$.  Note that our
Bayesian calculations focus on one particular realization from
the universe of possible realizations.  The global
proposition $D$ is the product of the propositions $D_i$ concerning the line
detections and nondetections in individual bursts.  Thus $D_i$ states that in
the $i$th burst certain line types were detected, and all others were not
detected.  For example, with two line types
$D_i=L_i (\epsvec_1) \bar L_i (\epsvec_2)$ indicates that in the $i$th burst
line type 1 was detected and line type 2 was not.

If the detections of different line types are not coupled then
the probability of observing $D$ is just the product of the probabilities of
each detection or nondetection as given by eqn.~(2).  The line types would be
coupled if the presence of lines of different types were correlated or if line
types could be confused; neither possibility is considered here.  For bursts
where $n_d$ lines are detected the probability for the data given $f$ (the
likelihood for $f$) is
\begin{eqnarray}
p \left(D_i \,|\, fHI\right)
&=& \prod_{\sigma=1}^{n_d} \left[ \alpha_{i\sigma} f_\sigma +
   \beta_{i\sigma} (1-f_\sigma) \right]
   \prod_{\sigma=n_d+1}^{n_t} \left[ \left(1-\alpha_{i\sigma}\right)
   f_\sigma + \left(1-\beta_{i\sigma}\right) (1-f_\sigma) \right] \\
&=& \prod_{\sigma=1}^{n_d} {{\alpha_{i\sigma} f_\sigma +
   \beta_{i\sigma} (1-f_\sigma)} \over
   {\left(1-\alpha_{i\sigma}\right)
   f_\sigma + \left(1-\beta_{i\sigma}\right) (1-f_\sigma)}}
   \prod_{\sigma=1}^{n_t} \left[ \left(1-\alpha_{i\sigma}\right)
   f_\sigma + \left(1-\beta_{i\sigma}\right) (1-f_\sigma) \right]
   \nonumber
\end{eqnarray}
where for clarity we number the detected line types first; a more complicated
indexing is necessary when different bursts with line detections are
considered.  The second
formulation in eqn.~(3) is more compact and leads to useful limits.  Note that
in eqn.~(3) the probability of the observed outcome $D_i$ is calculated for
all possible combinations of the presence and absence of the line types;
terms with a factor $f_\sigma$ assume the $\sigma$th line type is present
and those with $(1-f_\sigma)$ assume the line is absent.

For an ensemble of $N_G$ \Ginga\ and $N_B$ BATSE bursts which have been
searched the likelihood, the probability for the observed
realization, is
\begin{equation}
p\left(D \,|\, fHI\right) =
   \prod_{k=1}^{N_G} p\left(D_k \,|\,  fHI\right)
   \prod_{m=1}^{N_B} p\left(D_m \,|\, fHI\right)
\end{equation}
which is valid even when the line types are coupled (i.e., if eqns.~[1-3] are
not valid).

We write eqn. (4) one line type at a time for the current case of $n_G=2$ and
$n_B=0$ where we assume that there are only two line types.  For clarity we
place the line detection in the first \Ginga\ burst; this equation can be
extended easily to the second line type by reversing the definitions of the
first and second bursts.  The line frequency $f$ and detection probabilities
$\alpha$ and $\beta$ now refer to the single line
type under consideration, and will have different values for each line
type.  We assume that BATSE and \Ginga\ observe the same populations
of strong bursts and therefore their line frequencies should be the
same, but for the
purposes of the analysis below we write the likelihood in terms of separate
line frequencies for each detector
\begin{eqnarray}
p(D \,|\, f_G f_B I) &=&
   \left(\alpha_1 f_G +\beta_1\left(1-f_G \right)\right)
   \prod_{k=2}^{N_G} \left(1-\alpha_k f_G -\beta_k\left(1-f_G \right)\right)
   \quad \times \\
   && \prod_{m=1}^{N_B} \left(1-\alpha_m f_B -\beta_m\left(1-f_B \right)
   \right) \nonumber
\end{eqnarray}
where line type indices have been suppressed.

To make more concrete the dependencies on the numbers of \Ginga\ and BATSE
bursts in which lines could be detected, we present below simplified
heuristic calculations in which we set $\alpha=1$; frequently we will also
set $\beta=0$.  The numbers of \Ginga\ and BATSE bursts must be reduced
accordingly to compensate for the bursts in which lines could not be
detected.  Empirically we find this approximation is reasonable for
values of $N_G$ and $N_B$ equal to the sums of the actual $\alpha_i$.

The observed absorption lines are undoubtably drawn from a continuous line
parameter space.  Thus $f(\epsvec)$ should actually be a function of a number
of continuous variables.  The likelihood function can be derived from the
discrete line type likelihood.  Let the discrete $\epsvec_\rho$ be the vector
of average parameter values over a cell within the continuous parameter volume
$\Delta \epsvec_\rho$.  Then $p(L(\epsvec_\rho) \,|\, fI)$ is the probability
a line will be found within $\Delta \epsvec_\rho$.  If $p(\epsvec_\rho)$ is the
line detection probability distribution (i.e., probability per unit parameter
volume) then eqn.~(2) becomes
\begin{equation}
p(L(\epsvec_\rho) \,|\, fI) = p(\epsvec_\rho) \Delta \epsvec_\rho
   = \alpha(\epsvec_\rho)f_\rho \Delta \epsvec_\rho +
   \beta(\epsvec_\rho)\Delta \epsvec_\rho
   \left(1-f_\rho \Delta \epsvec_\rho \right)
\end{equation}
where we recognize that the probability of finding a false positive is
proportional to the parameter volume $\Delta\epsvec_\rho$.  Next we let $n_t$
go to infinity as $\Delta \epsvec$ goes to zero; $p(L(\epsvec_\rho) \,|\, fI)$
becomes a differential in the limiting process.

The confusion of one line type with another is unavoidable as we
pass to the continuum limit:  rarely will our spectral fits find the
exact line parameters.
The discrete likelihood functions derived above (which can easily be
generalized for a continuous parameter space) are not directly
relevant.  We
therefore defer derivation of continuous likelihoods until continuous line
distributions are determined from a much larger number of line detections
or are proposed by theories of burst emission.

\subsection{Bayesian Formalism}

In the Bayesian formulation of statistics, our confidence in a hypothesis'
truth is expressed in terms of a probability (this is one of the major
foundations of Bayesian statistics).  Thus $p(H \,|\, DI)$ is the
posterior probability that hypothesis $H$ is true given the data $D$ and
information $I$.  By Bayes' Theorem (a basic relation among
probabilities---Loredo 1990)
\begin{equation}
p(H \,|\, DI) = {{p(H \,|\, I)\,p(D \,|\, HI)}\over{p(D \,|\, I)}} \quad .
\end{equation}
The probability $p(D \,|\, HI)$ is the likelihood for $H$, and is the quantity
from which ``frequentist'' statistical methods derive standard quantities
such as $\chi^2$.  The probability $p(D\,|\,I)$ is the global likelihood,
the probability for the realization $D$ under all possible hypotheses; this
factor acts as a normalization (since we use ratios of $p(H\,|\, DI)$, we
need not calculate $p(D\,|\,I)$).  Finally, $p(H \,|\, I)$ is the
prior probability that the hypothesis is true, and is therefore a
quantification of our expectations.  Probabilities with no dependence on $D$,
such as $p(H\,|\,I)$, occur frequently within the Bayesian methodology, and
are called ``priors.''  Priors for the current data set may be posterior
probabilities from the evaluation of a different experiment or observation.

We are concerned with the truth of $H$.  However, the likelihood
$p(D \,|\, f H I)$ in eqn.~(4) is a function of the unknown line frequency
distribution $f$.  While $f$ is intrinsically interesting, for hypothesis
evaluation the value of $f$ is necessary only to determine the more
fundamental $p(D\,|\,fHI)$, or in Bayesian terminology, $f$ is a ``nuisance''
parameter.  We eliminate $f$ by
the Bayesian process of marginalization:  integrating over all possible values,
weighted by our prior expectation for this parameter's likely values, that is,
by $f$'s prior.  Thus
\begin{equation}
p(D \,|\, HI) = \int df(\epsvec) \,p(f(\epsvec) \,|\,HI)\,
   p(D \,|\, fHI)
\end{equation}
where the integration is over each line type.

To compare the relative probabilities that hypotheses $H_0$ and $H_x$
are true, we construct the posterior odds ratio
\begin{equation}
O_H = {{p(H_0 \,|\, DI)}\over {p(H_x \,|\, DI)}}
   = {{p(H_0 \,|\, I)\,p(D \,|\, H_0 I)}\over
   {p(H_x \,|\, I)\,p(D \,|\, H_x I)}}
   = {{p(H_0 \,|\, I)}\over {p(H_x \,|\, I)}} \,
   {{\int df \, p(f \,|\, H_0 I)\, p(D \,|\, fH_0 I)}\over
   {\int df \,p(f \,|\, H_x I)\, p(D \,|\, fH_x I)}} \, .
\end{equation}
This is the basic equation.  Note that we do not have to calculate
$p(D \,|\,I)$.  The likelihood ratio $p(D \,|\, H_0 I)/p(D \,|\, H_x I)$,
often called the Bayes factor $B$, can be calculated.  With the current
pattern of detections and nondetections, the
Bayes factor will usually
favor the hypothesis that the \Ginga\ and BATSE observations are inconsistent,
or that our understanding of these instruments is faulty.  On the other hand,
the factor $p(H_0 \,|\, I) / p(H_x \,|\, I)$, the prior odds ratio,
is an expression of our prior expectations of the relative truth of
each hypothesis.  As such, this factor will often be subjective; for example,
if $H_0$ states that the BATSE detectors function as expected (i.e., are
capable of detecting lines), and $H_x$
states that BATSE is unable to detect lines, then the
prior odds quantifies our confidence in the BATSE detectors.  Usually our
conclusion regarding BATSE-\Ginga\ consistency depends on our assessment of
the relative values of the Bayes factor and the ratio of the hypothesis priors
(i.e., whether the priors compensate for a Bayes factor unfavorable for
consistency).  The arbitrariness in $p(H_0 \,|\, I)/p(H_x \,|\, I)$ makes
explicit the subjectivity in deciding when an inconsistency exists.  Clearly
our threshold for accepting a conclusion consistent with our expectations is
less stringent than for a surprising conclusion.

We use the above likelihood function to estimate the line frequencies by
calculating the posterior distribution for $f$, $p(f(\epsvec) \,|\, DHI)$,
based on the observed realization $D$.  This distribution can be derived
from the BATSE, \Ginga\ or combined datasets (i.e., by using
different definitions of $D$).  By Bayes' Theorem
\begin{equation}
p(f(\epsvec) \,|\, DHI) = {{p(f(\epsvec)\,|\,HI)\,p(D\,|\, f(\epsvec)HI)}
   \over{p(D\,|\,HI)}}
   = {{p(f(\epsvec)\,|\,HI)\,p(D\,|\, f(\epsvec)HI)}\over
   { \int df(\epsvec) \,p(f(\epsvec)\,|\,HI)\,p(D\,|\, f(\epsvec)HI)}}
   \quad .
\end{equation}
The line frequency prior, $p(f(\epsvec)\,|\,HI)$, depends on $H$ and $I$.  For
a uniform prior, $p(f\,|\,HI)=1$, the posterior distribution for $f$ is
proportional to the probability of $D$ as a function of $f$.
\section{HYPOTHESES}
Using the Bayesian methodology presented above in \S 2 for detections in
two \Ginga\ and no BATSE bursts, we compare the hypothesis $H_0$ stating
that there is no inconsistency between the \Ginga\ and
BATSE results to specific hypotheses which contradict $H_0$.  In detail
the consistency hypothesis $H_0$ states that: the detection capabilities of
both instruments are understood; lines exist; and the detection threshold
has been set high enough to virtually eliminate false positives (i.e., we set
$\beta$=0).  Each line type has its own Bayes factor, and the overall odds
ratio is the product of the two Bayes factors and the ratio of hypothesis
priors (the prior odds).  In the following we
present the Bayes factor for the single detection
of a given line type in the \Ginga\ data, and none in the BATSE bursts.

First, define $H_1$ to be the hypothesis that BATSE is unable to detect lines,
even if they are present.  Thus we set BATSE's line detection probability
$\alpha$ to zero for hypothesis $H_1$.  Consequently the Bayes factor for the
comparison of $H_0$ to $H_1$ is
\begin{equation}
B_1 = {{\int df \, p(f \,|\, H_0I) \; \alpha_1 f \; \prod_{k=2}^{N_G}
   (1-\alpha_k f) \; \prod_{m=1}^{N_B} (1-\alpha_m f)}\over
{\int df \, p(f \,|\, H_1I) \;\alpha_1 f \;\prod_{k=2}^{N_G} (1-\alpha_k f)}}
\end{equation}
where $\alpha$ indexed with $k$ refers to \Ginga\ bursts and with $m$ to
BATSE bursts.  Since this is the Bayes factor for one line type, the line type
indices are suppressed. If for our heuristic calculation we set $\alpha=1$ for
hypothesis $H_0$ and reduce $N_G$ and $N_B$ to $N_G^\prime$ and $N_B^\prime$
(the strong bursts for which $\alpha\sim 1$), then
\begin{equation}
B_1 = {{\int df \, p(f \,|\, H_0I) \, f (1-f)^{N_G^\prime+N_B^\prime-1}}
   \over {\int df \, p(f \,|\, H_1I) \, f (1-f)^{N_G^\prime-1}}}
   = {{N_G^\prime(N_G^\prime+1)}\over
   {(N_G^\prime+N_B^\prime)(N_G^\prime+N_B^\prime+1)}} \quad ,
\end{equation}
where we used uniform line frequency priors $p(f \,|\, HI)=1$ in calculating
the last term; as will be discussed below (\S 4), these are formally correct
priors.  The analytic expressions
for the Bayes factor in the heuristic calculations use the Beta function
\begin{equation}
\int_0^1 df \, f^n (1-f)^{N-n} = {{n! (N-n)!}\over{(N+1)!}} \quad .
\end{equation}
Since $n$ is usually 0 or 1, and $N$ is fairly large, small values of $f$
dominate the integral.  Thus, if instead of a uniform prior from 0 to 1
we use a uniform prior from 0 to $f_{max}$, the dependence on $n$ and $N$
will be the same (to within $\sim25$\%), with a normalization factor (from
the prior) of $1/f_{max}$.  This normalization factor will appear in both the
denominator and numerator of eqn.~(12), and $B_1$ will change by very little.

Next, the hypothesis $H_2$ states there are no absorption lines.  Consequently
the reported \Ginga\ lines must all be false positives.  Thus the prior
$p(f \,|\, H_2 I) = \delta (f)$ and the false positive probability $\beta$
must be nonzero (and assumed constant) for $H_2$:
\begin{equation}
B_2 = {{\int df \, p(f \,|\, H_0I) \; \alpha_1 f \; \prod_{k=2}^{N_G}
   (1-\alpha_k f) \; \prod_{m=1}^{N_B} (1-\alpha_m f)}\over
{\beta_1 \;\prod_{k=2}^{N_G} (1-\beta_k) \; \prod_{m=1}^{N_B} (1-\beta_m)}}
\end{equation}
For our heuristic calculation we set $\alpha=1$ for $H_0$ and let $\beta$
be nonzero for $H_2$:
\begin{equation}
B_2 = {{\int df \, p(f \,|\, H_0I) \, f (1-f)^{N_G^\prime+N_B^\prime-1}}
   \over {\beta (1-\beta)^{N_G^\prime+N_B^\prime-1}}}
   = {1\over{\beta (1-\beta)^{N_G^\prime+N_B^\prime-1}
   (N_G^\prime+N_B^\prime)(N_G^\prime+N_B^\prime+1)}} \quad ,
\end{equation}
where again we evaluated the integral using a uniform line frequency prior.
If we cut off the prior at $f_{max}$, then $B_2$ increases by a
factor of $1/f_{max}$ (the uniform prior only appears in the numerator).
This Bayes factor can be minimized by maximizing
$\beta (1-\beta)^{N_G^\prime+N_B^\prime-1}$:
\begin{eqnarray}
\beta_{max} &=& {1\over{N_G^\prime+N_B^\prime}} \\
B_2 &=& {1\over{N_G^\prime+N_B^\prime+1}} \left({{N_G^\prime+N_B^\prime}
   \over{N_G^\prime+N_B^\prime-1}}\right)^{N_G^\prime+N_B^\prime-1} \quad .
\end{eqnarray}

Finally we use as a generalized inconsistency hypothesis $H_3$ the supposition
that the \Ginga\ and BATSE bursts are characterized by different line
frequencies.  If $H_3$ is favored we would not believe there actually are
different line frequencies, but instead would conclude the instruments are not
well understood and the line detectability probabilities are incorrect.  Note
that an error in the line detectability calculations, which can be modeled by
hypothesis $H_3$, need not imply that BATSE is unable to detect lines
(hypothesis $H_1$).  Differences between an instrument's true and calculated
line detection capabilities can indeed be modeled by changes in the line
frequency.  The Bayes factor is
\begin{equation}
B_3 = {{\int df \, p(f \,|\, H_0I) \; \alpha_1 f \; \prod_{k=2}^{N_G}
   (1-\alpha_k f) \;\prod_{m=1}^{N_B} (1-\alpha_m f)}\over
   {\int\!\!\int df_G df_B \; p(f_G \,|\, H_3I)p(f_B \,|\, H_3I) \;
   \alpha_1 f_G \; \prod_{k=2}^{N_G} (1-\alpha_k f_G) \; \prod_{m=1}^{N_B}
   (1-\alpha_m f_B)}}
\end{equation}
where $\alpha$ indexed with $k$ refers to \Ginga\ bursts and with $m$ to BATSE
bursts.  Note that the integral in the numerator can be viewed as the integral
in the denominator with an extra factor of $\delta(f_G-f_B)$ in the integrand.
The double integral over $f_G$ and $f_B$ for $H_3$ in this equation includes
$f_B=f_G$, but the fraction of the $f_G-f_B$ plane where $f_B = f_G$ is
infinitesimal, and therefore the case $f_B = f_G$ is given no weight in the
integral.  For our heuristic calculation we set $\alpha=1$ and
reduce $N_G$ and $N_B$ to $N_G^\prime$ and $N_B^\prime$
\begin{eqnarray}
B_3 &=& {{\int df \, p(f \,|\, H_0I) \, f (1-f)^{N_G^\prime+N_B^\prime-1}}
   \over {\int df_G \; p(f_G \,|\, H_3I) \, f_G (1-f_G)^{N_G^\prime-1} \;
   \int df_B \; p(f_B \,|\, H_3I) \, (1-f_B)^{N_B^\prime}}} \\
   &=& {{N_G^\prime(N_G^\prime+1)(N_B^\prime+1)}\over
   {(N_G^\prime+N_B^\prime)(N_G^\prime+N_B^\prime+1)}} \quad , \nonumber
\end{eqnarray}
where the last expression was calculated with uniform line frequency priors.
If the uniform prior extends only to $f_{max}$ then $B_3$ decreases by a
factor of $f_{max}$ since the prior occurs twice in the denominator and only
once in the numerator.

$B_3$ can be used to assess whether the \Ginga\ data increased our knowledge
of the line frequency given the BATSE results; the appropriate hypothesis
priors are required to answer this different question.  In this case $B_3$
evaluates the BATSE data alone using two different priors for $f$.  The
numerator uses a prior based on the \Ginga\ data (the integral over $f_G$ in
the denominator normalizes this prior) while the denominator uses a uniform
prior (the integral over $f_B$).  Here the posterior for $f$ from the \Ginga\
data is used as a prior for the BATSE observations.
\section{ILLUSTRATIVE CALCULATION}
A detailed calculation using the detection probabilities $\alpha$ and false
positive probabilities $\beta$ for each \Ginga\ and BATSE burst will be
presented in a subsequent paper in this series.  Here we present calculations
using the heuristic Bayes factors (eqns. [12], [17] and [19]).  The
illustrative set of $N_G^\prime$ and $N_B^\prime$ for the GB880205 and
GB870303 line types presented in Table~1 are based on more complete
calculations using preliminary values of $\alpha$ and $\beta$.  Therefore by
considering the Bayes factors presented here we can reach tentative conclusions
which will most likely remain valid after our more detailed calculation.  The
value of $N_B^\prime$ is surprisingly small given the large number of BATSE
bursts which have been searched:  most bursts
were not strong enough for \Ginga-like
spectral features to be detectable.  In addition, few BATSE spectra extend
below $\sim 20$~keV which would enable detection of GB870303-like lines, and
therefore we give $N_B^\prime$ a very small value for this line type.  The
\Ginga\ $N_G^\prime$ is based on the calculations of Fenimore et al. (1993).

For our primary analysis we assumed a uniform prior for the line frequencies,
except for hypothesis $H_2$ which states $f=0$; for all other hypotheses
$f$ can be any value between 0 and 1, or $p(f \,|\,HI)=1$.  Formally this
prior must utilize information prior to the \Ginga\ detector.  As discussed
in the Introduction, there is insufficient pre-\Ginga\
information to calculate a line frequency, and we use the least informative
line frequency prior (i.e., the prior with the least information content).
Table~1 lists the Bayes factor for each set of hypothesis comparisons using
the lines in GB880205 alone, the line in GB870303, and both line sets together
(the column labeled ``Joint'').  Note that the
posterior odds (eqn.~[9]), which
indicates the favored hypothesis, is the product of the Bayes factor (the
ratio of the likelihoods for each hypothesis) and prior odds (the
ratio of hypothesis priors) quantifying our expectations and knowledge before
the data were obtained.  Table~1 also presents the prior odds for each
hypothesis comparison (as discussed below), and the resulting posterior
odds.

Because the data appear discrepant with two \Ginga\ and no BATSE detections,
we might intuitively expect Bayes factors less than 1, favoring the specific
hypotheses regarding instrumental deficiencies ($H_1$---BATSE is unable to
detect lines---and $H_2$---absorption lines do not exist and therefore the
\Ginga\ detections must be spurious).  On the other hand, the prior odds
(the ratio of the hypothesis priors)
are greater than~1, favoring our assertion that the instruments are
understood.  Based on prelaunch calibration tests and on-orbit performance and
observations (e.g., the Her X-1 pulsar spectrum---Briggs et al. 1994) we are
confident that BATSE could detect lines if present (Teegarden et al. 1993;
Band et al. 1993a; Palmer et al. 1994a), and therefore we assign a high value
(e.g., $\sim100$) to the prior odds $p(H_0 \,|\, I)/p(H_1 \,|\, I)$.

The prior for hypothesis $H_2$ consists of the product of priors for both
line nonexistence and spurious Ginga detections.  Although line nonexistence
is the fundamental statement of $H_2$, a necessary consequence is that the
claimed detections are false positives; priors are required for both
propositions.  Before the report of the \Ginga\ lines the confidence in the
existence of absorption lines in GRB spectra
was not very high, and therefore formally our prior for line existence, which
should be based on pre-\Ginga\ information, is not large.  On the other
hand, an unrealistically large value of the probability of spurious detections
$\beta$ minimizes the Bayes factor for the $H_0$ to $H_2$ comparison; the
\Ginga\ team has studied their claimed detections and is confident they are
not false positives (E.~Fenimore 1993, private communication; C.~Graziani 1993,
private communication).  The detection threshold has been set high enough
to make the probability of a statistical false positive very small, and the
\Ginga\ instrument team worked hard to eliminate systematic effects which
could produce a spurious detection.  Note that a systematic effect could
increase the false positive probability significantly for all bursts.
Therefore, based on the expectations both that lines exist and that the false
detection probability is low, we assign a high value (e.g., $\sim$100) to the
prior odds of $H_0$ relative to $H_2$.

The Bayes factor $B_3$ is surprisingly close to 1 for the comparison of
consistency ($H_0$) and generalized inconsistency ($H_3$).  Figure~1 explores
the dependence of $B_3$ on $N_B^\prime$ for $N_G^\prime=10$ assuming there are
no line detections in the BATSE spectra; the Bayes factor for multiple line
types is the product of the single-detection Bayes factor.  This figure shows
the number of BATSE bursts without line-detections necessary to conclude there
is an inconsistency (the effective value of $N_G$ is $N_G^\prime \sim 10$).
As can be seen, for one line type $B_3 \propto 1/N_B^\prime$ when
$N_B^\prime\gg N_G^\prime$ (see also eqn.~[19]).  We assume we
understand our instruments and therefore assign prior odds favoring $H_0$
over $H_3$ (e.g., $\sim10$), although not by as large a factor of $H_0$
relative to $H_1$ or $H_2$ since the implications of $H_3$ are not as extreme
as these other two hypotheses.  Inaccuracies in our line detectability
calculations, which are more likely than BATSE's total inability to detect
lines ($H_1$), can be modeled as differences in the line frequencies.
Figure~2 shows $B_3$ for different values of $N_G^\prime$; $B_3$ increases
with $N_G^\prime$ since the likely line frequency from the \Ginga\ data alone
decreases.  We conclude from these figures that an order of magnitude more
strong BATSE bursts without a line detection are necessary for the odds
ratio to fall below unity, that is, for hypothesis $H_3$ to be favored.

The surprisingly large value of the Bayes factor comparing $H_0$ and $H_3$
results from the structure of the space of the likelihood function
$p(D \,|\, f_G f_B I)$ as a function of the \Ginga\ and BATSE line frequencies
$f_G$ and $f_B$ (eqn.~[5]).  The line frequencies are marginalized by
integrating over this space.  Figure~3 shows this space with logarithmic
contours for the likelihood with $N_G^\prime=10$ and $N_B^\prime=35$; while
this example is based on the values for GB880205 in Table~1, it is meant to be
illustrative.  Under
the hypothesis $H_0$ that there is a single line frequency $f=f_G=f_B$ the
line frequency is marginalized by integrating along the diagonal.  On the
other hand, for $f_G \ne f_B$ assumed by $H_3$ the integration is over the
entire region.  The peak value of the likelihood is not on the line $f_G=f_B$,
yet the average along this line is comparable to the average over the
entire region!

Figure~4 gives the distribution for likely values of the line frequency
$p(f \,|\, DH_0I)$ from eqn.~(10) for the \Ginga\ and BATSE datasets alone,
and for the joint dataset.  Note that the abscissa is the logarithm of the line
frequency, and therefore the areas under the curves are not proportional to
the probability assigned these regions.
As can be seen, there is a substantial overlap
between the line frequency distributions for each instrument's data alone.
Indeed, the distribution for $f$ from the joint dataset is the (normalized)
product of these two distributions.  For a range of $f$ values
the \Ginga\ detections can be a statistical fluctuation up and the absence
of BATSE detections a fluctuation down.  This explains the larger than
expected value of $p(D \,|\, H_0 I)$, the probability of obtaining the data
assuming consistency $H_0$.  On the other hand, we find a small value for
the probability of obtaining the BATSE results alone---the right hand
factor of $p(D \,|\, H_3 I)$,  the likelihood for $H_3$ (the
denominator of $B_3$).  With a uniform prior for the line frequency the
probability of detecting lines in $n_B$ bursts out of $N_B^\prime$ searched
is $1/(N_B^\prime+1)$, independent of $n_B$.  Therefore finding no lines in
the BATSE data is only one of many equally likely results, hence the small
value of its occurrence.

As was discussed above, we have been using uniform line frequency priors
between 0 and 1, $p(f \,|\, I)=1$, because we cannot determine
dependable line occurrence rates from the pre-\Ginga\ reports of line
detections.  Although there are many problems in assessing the line frequency
in the KONUS bursts, we can naively use a uniform prior to $f_{max}=0.2$
(the line frequency KONUS reported).  The resulting Bayes factors are provided
in Table~1.  As can be seen $B_1$ changes by less than a factor of 2, while
$B_2$ increases and $B_3$ decreases by factors of order $1/f_{max}\sim 5$.
However, our basic conclusions are unaffected.

\section{``FREQUENTIST'' ANALYSIS}
Before adopting the Bayesian methodology presented here, we defined the
consistency statistic as the probability $p(n_G \ge 2, n_B=0 \,|\, H_0 I)$
that \Ginga\ would detect 2 or more lines, and BATSE none (Band et al. 1993c).
This is the region in the space of all possible realizations where the
observations would appear to be at least as discrepant as the current
detections and nondetections.  A small value was understood to indicate
inconsistency between the \Ginga\
and BATSE results.  Frequentist calculations consider how likely the data
are for a given hypothesis.  However, the probability of obtaining the
observed data may be vanishingly small if there are many possible outcomes
(for example, observing a particular value of a continuous variable), and
therefore the probability is calculated for a region bounded by the
observations.
By working with only one hypothesis, we do not know whether the observations
are any more likely for any reasonable alternative to that hypothesis.
On the other hand, Bayesian statistics compares hypotheses using the
probabilities of obtaining the observed outcome under the hypotheses without
regard for the magnitude of these probabilities.  However, our lack of
imagination in devising reasonable alternative hypotheses may lull us into
complacency if we find odds ratios favoring the null hypothesis (here the
consistency hypothesis $H_0$) over unlikely hypotheses.
It is therefore instructive to consider our frequentist consistency measure.

Since the \Ginga\ lines are actually single detections of two very different
line types, we calculate the product of the probabilities of one or more
detections of two types.  For a single line type this probability is
\begin{equation}
P(n_G \ge 1, n_B=0 \,|\, f, N_G,N_B) = \prod_{m=1}^{N_B}
   (1-\alpha_m f-\beta_m(1-f)) \left(1- \prod_{k=1}^{N_G}
   \left(1-\alpha_k f-\beta_k\left(1-f\right)\right) \right) \quad .
\end{equation}
As with the Bayesian analysis we simplify this expression to see the
dependencies on the number of bursts.  Thus we set $\alpha=1$ and $\beta=0$,
and use the effective number of bursts $N_B^\prime$ and $N_G^\prime$:
\begin{equation}
P(n_G \ge 1, n_B=0 \,|\, f,N_G^\prime,N_B^\prime) = (1-f)^{N_B^\prime}
   \left(1-(1-f)^{N_G^\prime}\right) \quad .
\end{equation}
These expressions are functions of the unknown line frequency $f$.  By
maximizing this probability with respect to $f$ we establish an upper limit
for consistency.  The line frequency which maximizes the probability
in eqn.~(21) is $\hat f=1-(N_B^\prime/(N_B^\prime+N_G^\prime))^{1/N_G^\prime}$,
giving
\begin{equation}
P_{max} (n_G \ge 1, n_B=0\,|\,\hat f, N_G^\prime,N_B^\prime) =
   \left({{N_B^\prime}
   \over{N_G^\prime+N_B^\prime}}\right)^{N_B/N_G} \left({{N_G^\prime}\over
   {N_G^\prime+N_B^\prime}}\right) \quad .
\end{equation}
Table~2 lists this probability evaluated for the values of the $N_G^\prime$
and $N_B^\prime$ in Table~1.
As can be seen, maximizing the probability with respect to $f$ gives an upper
limit of 3\% that \Ginga\ and BATSE will appear as discrepant if lines exist
and the detectors function as understood.

An alternative consistency measure is the probability that all the detections
would be in the \Ginga\ bursts given a set number of detections, the product of
$P(n_G=1,n_B=0 \,|\, n_G+n_B=1, N_G,N_B)$ for each line type (Palmer et al.
1994a).  This probability for one line type, assuming $\alpha=0$ or 1, is
\begin{equation}
P(n_G=1,n_B=0 \,|\, n_G+n_B=1, N_G^\prime,N_B^\prime) =
   {{N_G^\prime}\over{N_G^\prime+N_B^\prime}} \quad .
\end{equation}
Table~2 presents this consistency measure evaluated for our illustrative
example; there is a 13\% probability that both detections would be in the
\Ginga\ bursts, which would hardly be considered a discrepancy.  This
measure is more favorable for consistency than the previous one (eqn.~[22])
because it assumes there is only a single detection of a given line type,
thereby restricting the space from which our observed result is drawn.

Note that these two frequentist consistency measures test whether the
detections of a
given line type in one instrument but not another constitutes a discrepancy
between these two instruments, but not whether finding all the detections
in the same instrument is a discrepancy.  Indeed, these two measures would
have been smaller had the GB870303 line
been detected by BATSE and not by \Ginga!  Yet
in that case we would not worry about a discrepancy between BATSE and \Ginga.
It is only when we compare consistency and inconsistency hypotheses that we
test explicitly whether the instruments are discrepant.
\section{DISCUSSION}
The primary purpose of this paper is the development of a methodology
to compare the \Ginga\ and BATSE observations; the calculation
of the actual detection and false positive probabilities (the $\alpha$
and $\beta$ quantities in the above equations) will be presented later in
this series.  Nonetheless, the example we used is a reasonable approximation
to the observations, and its analysis indicates the likely results of a more
accurate calculation.  The frequentist consistency statistic
$P (n_G \ge 2,n_B=0 \,|\, f, N_G,N_B)$ indicates that the detection
of at least two line features in the \Ginga\ bursts, and none in the BATSE
bursts, is fairly improbable (the probability has an upper limit of
$\sim 3$\%), but not unlikely enough to conclude there is
an inconsistency.  In addition, the probability that the two detections would
both be found in the \Ginga\ bursts is 13\%, which is not small enough to
indicate there is a discrepancy.  From our Bayesian odds ratios we infer that
the quantitative analysis of the data (represented by the Bayes factors) is
insufficient to shake our confidence in our understanding of the two
detectors.  Conversely, the Bayes factors do not prove conclusively that
there is not a discrepancy; the data do not rule out a serious deficiency in
the capabilities of either instrument, or in the analysis and interpretation
of their observations.  Therefore we continue to test BATSE's line-detecting
capability, and to study issues such as the false positive probability.

Bayesian inference has been faulted for the uncertainty as to the correct
prior, and indeed in the calculations we present in Tables~1 we use two
different priors for the line frequency.  However, it should be noted that
the basic conclusions are the same.
As stated above in \S 4, the uniform prior between $f=0$ and~1 is formally
correct in not using the BATSE or \Ginga\ data, and is also the most
conservative in not attempting a quantitative estimate of the line frequency
from the Konus data.  Therefore, the determination of the line
frequency prior does
not introduce any ambiguity into our conclusions.  Note that our conclusions
do depend on the hypothesis priors, the quantification of the confidence in
the analysis of the \Ginga\ and BATSE spectra.

BATSE observes strong bursts within which lines are detectable at a low enough
rate that it is unlikely the statistical analysis presented here will lead us
to conclude there is an inconsistency in the near future.  Figures~1 and~2
indicate that many more than 100 strong BATSE bursts would be necessary to
conclude the \Ginga\ and BATSE bursts are characterized by different line
frequencies (or alternatively, the line detection rate is very different than
calculated).  Similarly, a much larger number of BATSE bursts is necessary
for the Bayesian odds ratios comparing consistency and specific instrumental
deficiency hypotheses to convince us there is a discrepancy.  Therefore, in
the near term the continued absence of BATSE detections will merely lower our
estimate of the line frequency.

A major but unavoidable deficiency of our analysis is that we approximate
the continuous line distribution by the small number of lines detected.  Our
comparison of the two datasets is
necessarily plagued by uncertainty concerning the line distributions.  Of
course, sufficient line detections to determine these distributions would
prove definitively the existence of absorption lines in burst spectra, making
irrelevant the statistical analysis of the possible discrepancy between
\Ginga\ and BATSE, and permitting the more satisfying study of an important
burst phenomenon.
\section{SUMMARY}
We adopted a new Bayesian methodology to determine whether the two \Ginga\
detections of absorption lines and the absence of any BATSE detections are
inconsistent.  This methodology permits us to compare specific hypotheses
through an odds ratio which is the product of a quantitative Bayes factor,
the ratio of the probabilities of obtaining the
observations given the hypotheses, and a more subjective factor quantifying
our prior expectations.

The definitive application of this methodology to the BATSE and \Ginga\ data
requires detailed information on the bursts and line detectability, and will
be presented later in this series of papers.  However, we can draw tentative
conclusions based on an approximate calculation.  We find that the Bayes
factors favor hypotheses that the understanding of the \Ginga\ and BATSE
detectors are defficient, but not by large enough factors to exceed
our confidence in the understanding of these instruments.  Similarly,
the Bayes factor is inconclusive for a comparison of the hypotheses that the
\Ginga\ and BATSE bursts are characterized by the same or different line
frequencies.  Thus given the tests to which the \Ginga\ and BATSE instruments
have been subjected, our Bayesian methodology leads us to conclude that the
two instruments are not discrepant.  In addition, the non-Bayesian consistency
probabilities are not small enough to lead us to conclude there is an
inconsistency.

\acknowledgments{We thank the referee, Tom Loredo, for his
insightful (and copious) comments
which have improved this paper both in content and clarity.  The BATSE
instrument team effort at UCSD is supported by NASA contract NAS8-36081.}

\vfill\eject

\null\vskip2truein
\begin{table}
\caption{Illustrative Bayesian Calculation}
\begin{tabular}{l c c c c c}
\tableline
\tableline
& GB880205 & GB870303 & Joint & & \\
\tableline
$N_G^\prime$ & 10 & 15 & && \\
$N_B^\prime$ & 35 & 10 & && \\
$\beta_{max}$ & 1/45 & 1/25 & 1/35 && \\
\tableline
& \multicolumn{3}{c}{Bayes Factors} & Prior Odds & Posterior Odds \\
\tableline
\multicolumn{6}{l}{Uniform Line Frequency Prior} \\
$H_0:H_1$ & 0.0531 & 0.369 & 0.0196 & $\sim 100$ & $\sim 2$ \\
$H_0:H_2$ & ${{4.83\times10^{-4}}\over{\beta(1-\beta)^{44}}}$
   & ${{1.54\times10^{-3}}\over{\beta(1-\beta)^{24}}}$
   & ${{7.43\times10^{-7}}\over{\beta^{2}(1-\beta)^{68}}}$ & $\sim 100$ &
   $\sim {{7\times10^{-5}}\over{\beta^{2}(1-\beta)^{68}}}$ \\
$H_0:H_2$, $\beta=\beta_{max}$ & 0.0584 & 0.102 & 0.00653 & $\sim 100$ &
   $\sim 0.7$\\
$H_0:H_3$ & 1.91 & 4.06 & 7.75 & $\sim 10$ & $\sim 80$ \\
\tableline
\multicolumn{6}{l}{Uniform Line Frequency Prior 0-0.2} \\
$H_0:H_1$ & 0.0784 & 0.420 & 0.0329 & $\sim100$ & $\sim 3$ \\
$H_0:H_2$ & ${{2.41\times10^{-3}}\over{\beta(1-\beta)^{44}}}$
   & ${{7.52\times10^{-3}}\over{\beta(1-\beta)^{24}}}$
   & ${{1.81\times10^{-5}}\over{\beta^{2}(1-\beta)^{68}}}$
   & $\sim100$ & ${{\sim 2\times10^{-3}}\over{\beta^{2}(1-\beta)^{68}}}$ \\
$H_0:H_2$, $\beta=\beta_{max}$ & 0.292 & 0.501 & 0.160 & $\sim100$
   & $\sim 20$ \\
$H_0:H_3$ & 0.56 & 1.01 & 0.57 & $\sim10$ & $\sim 6$ \\
\end{tabular}
\tablecomments{
\newline
$N_G^\prime$---number of \Ginga\ bursts in which lines are detectable
\newline
$N_B^\prime$---number of BATSE bursts in which lines are detectable
\newline
$\beta_{max}$---the false positive probability which minimizes $B_2$
\newline
$H_0:H_x$---Comparison of hypotheses $H_0$ and $H_x$, where:
\newline\hskip 25truept
$H_0$---\Ginga\ and BATSE are consistent
\newline\hskip 25truept
$H_1$---BATSE is unable to detect lines
\newline\hskip 25truept
$H_2$---lines do not exist and thus the \Ginga\ detections are spurious
\newline\hskip 25truept
$H_3$---different line frequencies characterize the BATSE and \Ginga\ bursts}
\end{table}
\vfill\eject
\null\vskip2truein
\begin{table}
\caption{Illustrative Frequentist Consistency Statistics}
\begin{tabular}{l c c c }
\tableline
\tableline
& GB880205 & GB870303 & Joint \\
\tableline
$N_G^\prime$ & 10 & 15 & \\
$N_B^\prime$ & 35 & 10 & \\
\tableline
$P_{max}(n_G\ge 1,n_B=0 \,|\, \hat f, N_G^\prime N_B^\prime)$
& $9.2\times10^{-2}$ & $3.3\times 10^{-1}$ & $3.0\times10^{-2}$ \\
$P(n_G=1,n_B=0 \,|\, n_G+n_B=1, N_G^\prime N_B^\prime)$
& $2.2\times 10^{-1}$ & $6.0\times 10^{-1}$ & $1.3\times 10^{-1}$ \\
\tableline
\end{tabular}
\end{table}
\vfill\eject
\centerline{\bf Figures}

Figure 1.  Bayes factor $B_3$ vs. number of BATSE bursts $N_B^\prime$ without
a line detection for a single detection in $N_G^\prime=10$ \Ginga\ bursts.
Shown are curves for one
(solid), two (short dashes) and three (long dashes) different line types.  The
Bayes factor compares the hypothesis $H_0$ that \Ginga\ and BATSE are
consistent to the generalized inconsistency hypothesis $H_3$ that the \Ginga\
and BATSE bursts are characterized by different line frequencies.  A uniform
prior probability was used for the line frequencies.  Lines are assumed to
be detectable if present.

Figure 2.  Bayes factor $B_3$ vs. number of BATSE bursts $N_B^\prime$ for a
single detection in $N_G^\prime=5$ (solid curve), 10 (short dashes), 15 (long
dashes) and 20 (dot-dot-dash) \Ginga\ bursts.  The illustrative values used
in Table~1 use $N_G^\prime=10$ for GB880205 and $N_G^\prime=15$ for GB870303.

Figure 3.  The probability (eqn.~[5]) of one line detection in $N_G^\prime=10$
\Ginga\ bursts and no detections in $N_B^\prime=35$ BATSE bursts in $f_G-f_B$
space.  $f_G$ and $f_B$ are the \Ginga\ and BATSE line frequencies, allowed
to be different.  Logarithmic contours spaced factors of 100 apart are used;
the maximum occurs at $f_G=0.1$,
$f_B=0$.  The line frequency for $H_0$ is marginalized by integrating along
the diagonal $f_G=f_B$ while the \Ginga\ and BATSE line frequencies $f_G$ and
$f_B$ are marginalized for $H_3$ by integrating over the entire region.

Figure 4.  Normalized distributions of line frequencies for one \Ginga\
detection out of $N_G^\prime=10$ bursts, and no BATSE detections out of
$N_B^\prime=35$ bursts.  Shown are distributions based on the BATSE (long
dashes), \Ginga\ (solid curve) and combined (short dashes) datasets.  Note that
the abscissa is logarithmic, and therefore areas are not proportional to
the probabilities assigned to different regions.


\begin{references}

\reference
Band, D., Palmer, D. Teegarden,~B., Ford,~L., Briggs,~M., Matteson,~J.,
Schaefer,~B., Paciesas,~W., \& Pendleton,~G. 1993a, in {\it Contributed
Papers of the 23rd International Cosmic Ray Conference}, 1-105
%
\reference
Band, D., et al. 1993b, \apj, 413, 281
%
\reference
Band, D., Ford,~L., Palmer, D. Teegarden,~B., Briggs,~M., Matteson,~J.,
Schaefer,~B., Paciesas,~W., \& Pendleton,~G. 1993c, in {\it Contributed
Papers of the 23rd International Cosmic Ray Conference}, 1-120
%
\reference
Briggs, M., et al. 1994, in preparation
%
\reference
Fenimore, E. E., Schwarz, G., Lamb,~D.~Q., Freeman,~P., \& Murakami,~T. 1993,
in {\it Proc. of the Compton Symposium}, eds. M.~Friedlander, N.~Gehrels and
D.~J.~Macomb (AIP: New York), 917
%
\reference
Graziani, C., Fenimore, E. E., Murakami, T., Yoshida, A., Lamb,~D.~Q.,
Wang,~J.~C.~L., \& Loredo,~T.~J. 1992, in {\it Proceedings of the Taos
Workshop on Gamma-Ray Bursts}, ed. C.~Ho., R.~I.~Epstein, and E.~E.~Fenimore
(Cambridge U. Press:  Cambridge), 407
%
\reference
Hueter, G. J. 1987, PhD Thesis, UC San Diego
%
\reference
Loredo, T. J. 1990, in {\it Maximum Entropy and Bayesian Methods}, ed.
P.~Foug\`ere (Kluwer Academic Publishers: Dordrecht), 81
%
\reference
Martin, B. R. 1971, {\it Statistics for Physicists} (Academic Press: London)
%
\reference
Mazets, E. P. et al. 1981, Nature, 290, 378
%
\reference
Mazets, E. P., Golenetskii,~S.~V., Ilyinskii,~V.~N., Guryan,~Yu.~A.,
Aptekar,~R.~L., Panov,~V.~N., Sokolov,~I.~A., Sokolova,~Z.~Ya., \&
Kharitonova,~T.~V. 1982, \apss, 82, 261
%
\reference
Mazets, E. P., Golenetskii,~S.~V., Guryan,~Yu.~A., Aptekar,~R.~L.,
Ilyinskii,~V.~N., \& Panov,~V.~N. 1983, in {\it Positron-Electron Pairs
in Astrophysics, AIP Conf. Proc. 101}, eds. M.~L.~Burns, A.~K.~Harding,
and R.~Ramaty (AIP:  New York), 36
%
\reference
Meegan, C. A., Fishman, G. J., Wilson, R. B., Paciesas,~W.~S.,
Pendleton,~G.~N., Horack,~J.~M., Brock,~M.~N., \& Kouveliotou,~C. 1992,
Nature, 355, 143
%
\reference
Murakami, T., et al 1988, Nature, 335, 234
%
\reference
Palmer, D., et al. 1994a, in {\it Proc. of the Huntsville Gamma Ray Burst
Workshop}, eds. G.~Fishman, K.~Hurley and J. Brainerd (AIP:  New York),
in press
%
\reference
Palmer, D., et al. 1994b, in preparation
%
\reference
Teegarden, B. et al. 1993, in {\it Proc. of the Compton Symposium}, eds.
M.~Friedlander, N.~Gehrels and D.~J.~Macomb (AIP: New York), 860
%
\reference
Wang, J. C. L., et al. 1989, \prl, 63, 1550
%
\reference
Yoshida, A., Murakami, T., Nishimura,~J., Kondo,~I., \& Fenimore,~E.~E. 1991,
\pasj, 43, L69

\end{references}
\end{document}